\documentclass[aps,prc,twocolumn,nofootinbib,amsmath,showpacs,superscriptaddress,floatfix]{revtex4}
\usepackage{graphicx,epsfig,epstopdf,longtable}
\usepackage{mathptmx}
\usepackage[light,first]{draftcopy} 
\usepackage[pdftex]{hyperref}
\usepackage{wasysym}

\newcommand{\beqy}{\begin{eqnarray}}
\newcommand{\eeqy}{\end{eqnarray}}
\newcommand{\bmlet}{\begin{subequations}}
\newcommand{\emlet}{\end{subequations}}
\newcounter{saveeqn}

\def\gsimeq{\,\,\raise0.14em\hbox{$>$}\kern-0.76em\lower0.28em\hbox  
{$\sim$}\,\,}  
\def\lsimeq{\,\,\raise0.14em\hbox{$<$}\kern-0.76em\lower0.28em\hbox  
{$\sim$}\,\,}  

\begin{document}

\title{Scissors resonance in the quasi-continuum of Th, Pa and U isotopes}

\author{M.~Guttormsen}
\email{magne.guttormsen@fys.uio.no}
\affiliation{Department of Physics, University of Oslo, N-0316 Oslo, Norway}
\author{L.A.~Bernstein}
\affiliation{Lawrence Livermore National Laboratory, 7000 East Avenue, Livermore, CA 94550-9234, USA}
\author{A.~G{\"o}rgen}
\affiliation{Department of Physics, University of Oslo, N-0316 Oslo, Norway}
\author{B.~Jurado}
\affiliation{CENBG, CNRS/IN2P3, UniversitŽ Bordeaux, Chemin du Solarium B.P. 120, 33175 Gradignan, France}
\author{S.~Siem}
\affiliation{Department of Physics, University of Oslo, N-0316 Oslo, Norway}
\author{M.~Aiche}
\affiliation{CENBG, CNRS/IN2P3, UniversitŽ Bordeaux, Chemin du Solarium B.P. 120, 33175 Gradignan, France}
\author{Q.~Ducasse}
\affiliation{CENBG, CNRS/IN2P3, UniversitŽ Bordeaux, Chemin du Solarium B.P. 120, 33175 Gradignan, France}
\author{F.~Giacoppo}
\affiliation{Department of Physics, University of Oslo, N-0316 Oslo, Norway}
\author{F.~Gunsing}
\affiliation{CEA Saclay, DSM/Irfu/SPhN, F-91191 Gif-sur-Yvette Cedex, France}
\author{T.W.~Hagen}
\affiliation{Department of Physics, University of Oslo, N-0316 Oslo, Norway}
\author{A.C.~Larsen}
\affiliation{Department of Physics, University of Oslo, N-0316 Oslo, Norway}
\author{M.~Lebois}
\affiliation{Institut de Physique Nucleaire d'Orsay, Bat. 100, 15 rue G. Glemenceau, 91406 Orsay Cedex, France}
\author{B.~Leniau}
\affiliation{Institut de Physique Nucleaire d'Orsay, Bat. 100, 15 rue G. Glemenceau, 91406 Orsay Cedex, France}
\author{T.~Renstr{\o}m}
\affiliation{Department of Physics, University of Oslo, N-0316 Oslo, Norway}
\author{S.J.~Rose}
\affiliation{Department of Physics, University of Oslo, N-0316 Oslo, Norway}
\author{T.G.~Tornyi}
\affiliation{Department of Physics, University of Oslo, N-0316 Oslo, Norway}
\affiliation{Institute of Nuclear Research of the Hungarian Academy of Sciences (MTA Atomki), Debrecen, Hungary}
\author{G.M.~Tveten}
\affiliation{Department of Physics, University of Oslo, N-0316 Oslo, Norway}
\author{M.~Wiedeking}
\affiliation{iThemba LABS, P.O. Box 722, 7129 Somerset West, South Africa}
\author{J.N.~Wilson}
\affiliation{Institut de Physique Nucleaire d'Orsay, Bat. 100, 15 rue G. Glemenceau, 91406 Orsay Cedex, France}

\date{\today}

\begin{abstract}
The $\gamma$-ray strength function in the quasi-continuum has been measured for $^{231-233}$Th, $^{232,233}$Pa
and $^{237-239}$U using the Oslo method. 
All eight nuclei show a pronounced increase in $\gamma$ strength at $\omega_{\rm SR}\approx 2.4$ MeV,
which is interpreted as the low-energy $M1$ scissors resonance (SR).
The total strength is found to be $B_{\rm SR} = 9-11 \mu_{N}^{2}$ when integrated
over the  $1 - 4$ MeV $\gamma$-energy region. The SR displays a double-hump structure that is
theoretically not understood. Our results are compared with 
data from ($\gamma$, ${\gamma} ^{\prime}$) experiments and theoretical sum-rule estimates
for a nuclear rigid-body moment of inertia.
\end{abstract}


\maketitle

\section{Introduction}
\label{sec:int}
Atomic nuclei in the actinide region are unique from an astrophysics point of view, because
they are purely made in rapid neutron-capture processes in
explosive stellar environments~\cite{ar07}. Attempts have been made
to use the abundances of $^{232}$Th 
and $^{235,238}$U observed in the solar system (measured from meteoritic analyses) 
to estimate the age of the Galaxy, although these estimates are very uncertain and model-dependent.
Thorium has been observed in stars similar to the Sun, and also in older metal-poor stars~\cite{ar07}.

To predict the abundance of the actinides, one has to know the relevant reaction rates 
not only for the most long-lived nuclei, e.g. $^{232}$Th (14.05 Gy) and $^{238}$U (4.468 Gy), 
but also for the ones with extremely high neutron excess. Therefore, one must rely on 
calculations to estimate unknown cross sections where experimental data are lacking. 
This is not only relevant for the astrophysical nucleosynthesis~\cite{ar07,kaeppler2011},
but also for future and existing nuclear reactors~\cite{chadwick2011}.

Together with optical-model potentials, the nuclear level density and $\gamma$-ray strength 
function ($\gamma$SF) are crucial inputs for calculating neutron-induced 
reaction cross sections for neutron energies above the neutron-resonance region. 
These quantities provide information on the average properties of excited nuclei 
and are particularly applicable for describing gross features in the quasi-continuum region, 
where the number of levels is too high to measure individual states and their transitions. 
To ensure a reliable estimation of unknown cross sections, a detailed knowledge of 
both the level density and $\gamma$SF is vital.

An enhancement of the $\gamma$SF may boost the 
$\gamma$ decay relative to other decay branches such as particle emission or fission.
For the actinides, which have deformed shapes, the low-energy orbital $M1$ scissors resonance (SR)
contributes significantly to the $\gamma$-decay probability. 

The first geometrical description of the  
SR was given by Lo Iudice and Palumbo~\cite{iudice1978}.
Naively, 
the SR can be viewed as the proton and neutron clouds oscillating
against each other like scissor blades.
For deformed nuclei, Chen and Leander~\cite{chen} 
predicted strong $M1$ transitions between $\Delta \Omega = 1$ Nilsson
orbitals\footnote{The single-particle Nilsson orbitals are labeled by $\Omega^{\pi}\left[Nn_z\Lambda\right]$,
where $\Omega$ is the projection of the angular momentum vector $j$ on the nuclear symmetry axis.} 
originating from the same spherical state. These predictions were later 
supported by the observation of an enhancement at $E_{\gamma}\approx 2.2$ MeV
in the $\gamma$ spectra of the excited $^{161}$Dy nucleus~\cite{gutt84}.

Discrete scissors states built on the ground state can be populated in the
($\gamma$, ${\gamma} ^{\prime}$) and ($e, e^{\prime}$) reactions.
Here, the strength, spin and in some cases the parity of
the strongest scissors states in $^{232}$Th 
and $^{235,236,238}$U have been determined with typical summed strengths of
$B_{\rm SR} \sim 3-4\  \mu_N^2$~\cite{heil1988,margraf1990,yevetska2010,adekola2011},
where $\mu_N$ is the nucleon magneton.
Because such measurements rely on the
identification of single states in an energy region of
$10^{4} - 10^{5}$ levels per MeV, it is reasonable to believe that not all the
strength has been experimentally resolved.

Recently~\cite{heyde2011}, a review of several experiments 
and various models on the SR has been presented. The microscopic description of the SR is based on
single-particle couplings between orbitals of 
the same angular momentum $\ell$ and $j$. These proton and neutron two-quasiparticle
configurations contribute in a more or less
coherent way. Therefore,  the macroscopic picture of
oscillating scissors blades is rather oversimplified.
Recent quasiparticle random phase approximation (QRPA) calculations~\cite{heyde2011,kuliev2010} are generally in
agreement with the observed energies of the scissors states and strengths observed in
($\gamma$, ${\gamma} ^{\prime}$) and ($e, e^{\prime}$) reactions.

According to the generalized Brink hypothesis~\cite{brink}, the SR is not only built on the nuclear
ground state, but on all excited states in the nucleus.
The Oslo method, which is based on particle-$\gamma$ coincidences, makes it possible to explore
the decay of SR states in the quasi-continuum region. The method permits the extraction
of both level density and $\gamma$SF in one and the same experiment~\cite{Schiller00,Lars11}.
These measurements cover the rather unexplored $\gamma$- and excitation-energy region up
to the neutron binding energy (or the threshold for fission). 
Recently, the level densities of $^{231-233}$Th and $^{237-239}$U~\cite{nld2013}
and the $\gamma$SFs in $^{231-233}$Th and $^{232,233}$Pa~\cite{guttormsen2012} 
using this method have been reported.
 \begin{figure}[t]
 \begin{center}
 \includegraphics[clip,width=\columnwidth]{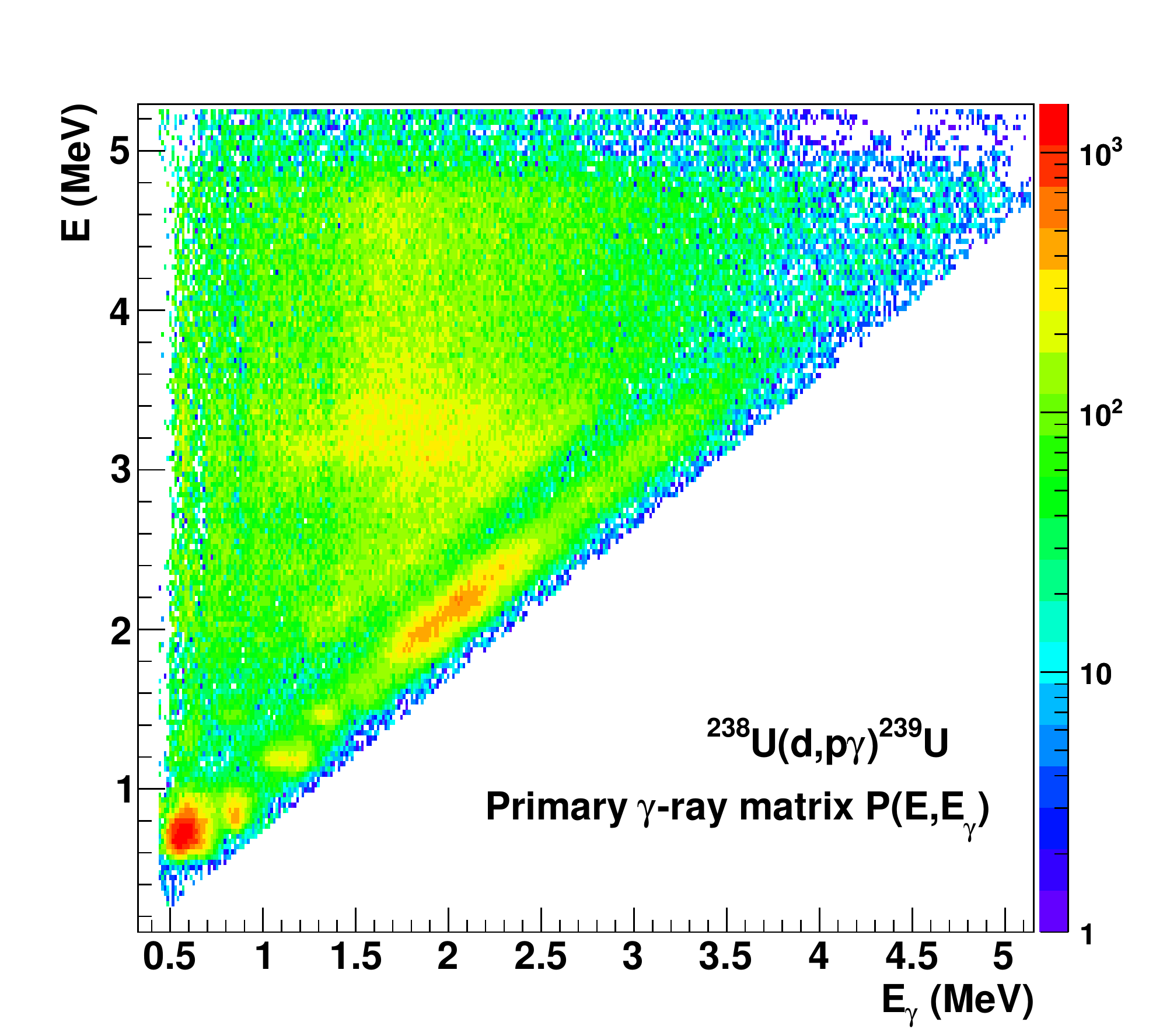}
 \caption{(Color online) First generation (primary) $\gamma$-ray matrix in $^{239}$U.
 At each excitation energy bin, $\gamma$ spectra can be projected out,
 giving the energy distribution of the first $\gamma$s from that excitation energy.
 The excitation energy ($E$) and $\gamma$ energy ($E_{\gamma}$) axis have dispersion 
 14.0~keV/ch and 32.4~keV/ch, respectively.}
 \label{fig:matrix}
 \end{center}
 \end{figure}

The main purpose of the present work is to make a comprehensive and systematic analysis  
of several actinides by exploiting nine reactions in total. The previous data
of the  $\gamma$SFs of $^{231-233}$Th and $^{232,233}$Pa~\cite{guttormsen2012} 
are reanalyzed and new experiments on $^{237-239}$U are presented.
In addition, the level densities of $^{232,233}$Pa are reported for the first time.

The structure of the manuscript is as follows. Section II describes the experimental techniques and methods,
and in Sect.~III the extraction and normalization of the $\gamma$SFs are discussed. 
In Sect.~IV the SRs are presented and extracted resonance parameters are given.
Section~V compares the data with previous results and models. Conclusions are drawn in Sect.~VI.

\section{Experiments}
\label{sec:exp}

The experiments with were performed with the MC-35
Scanditronix cyclotron at the Oslo Cyclotron Laboratory (OCL). 
The selfsupporting $^{232}$Th target (thickness 0.968 mg/cm$^2$) was bombarded with a 12-MeV deuteron 
and a 24-MeV $^{3}$He beam. The $^{238}$U target (thickness 0.260 mg/cm$^2$ and enrichment 99.7\%) had
a carbon backing (thickness 0.043 mg/cm$^2$) and was bombarded with a 15-MeV deuteron beam.
Particle-$\gamma$ coincidences were measured with the SiRi particle telescope and the 
CACTUS $\gamma$-detector system~\cite{siri,CACTUS}.

In order to reduce the intense elastically scattered projectiles on the detectors and exposure of deuteron break-up,
the 64 SiRi telescopes were placed in backward direction
covering eight angles from $\theta = 126^\circ$ to $140^\circ$
relative to the beam axis. These angles also give a broader and higher spin distribution
that are in better agreement with the real spin distribution of the nucleus. The front
and back detectors have thicknesses of $130$~$\mu$m and $1550$~$\mu$m, respectively. 
The CACTUS array consists of 28 collimated $5^{\prime\prime} \times 5^{\prime\prime}$ NaI(Tl) 
detectors with a total efficiency of $15.2$\% at $E_{\gamma} = 1.33$~MeV. 

The particle-$\gamma$ coincidences with time information were sorted event by event.
Gates were set on the 64 $\Delta$E-E matrices to select various particle types.
From the known charged-particle type and the kinematics of the reaction, the energies deposited in the telescopes 
can be translated to initial excitation energy $E$ in the residual nucleus. To avoid contamination from $\gamma$
rays emitted by the fission fragments, we consider only excitation energies below the fission barrier. 
For each energy bin $E$, 
the $\gamma$-spectra are unfolded~\cite{gutt6} using new NaI-response functions based on 
several in-beam $\gamma$ lines from excited
states in $^{13}$C, $^{16,17}$O, $^{28}$Si and $^{56,57}$Fe, where the relative detector
efficiency as function of $\gamma$ energy could be extracted in a reliable way.

An iterative subtraction technique was applied to separate out the 
first-generation (primary) $\gamma$ transitions from the total $\gamma$ 
cascade~\cite{Gut87}.  The  
technique is based on the assumption that the $\gamma$ distribution
is the same whether the levels were populated directly
by the nuclear reaction or by $\gamma$ decay from higher-lying states. 
This assumption is necessarily fulfilled when states have the same
relative probability to be populated by the two processes, since $\gamma$-branching 
ratios are properties of the levels themselves. If the excitation bins contain many 
levels, as is the case for the actinides, it is likely to find the same $\gamma$ distribution
independent of the method of population. Figure~\ref{fig:matrix} shows the final first-generation
$\gamma$-ray matrix $P(E,E_{\gamma})$ for the $^{238}$U($d, p\gamma$)$^{239}$U stripping reaction.
 \begin{figure}[t]
 \begin{center}
 \includegraphics[clip,width=\columnwidth]{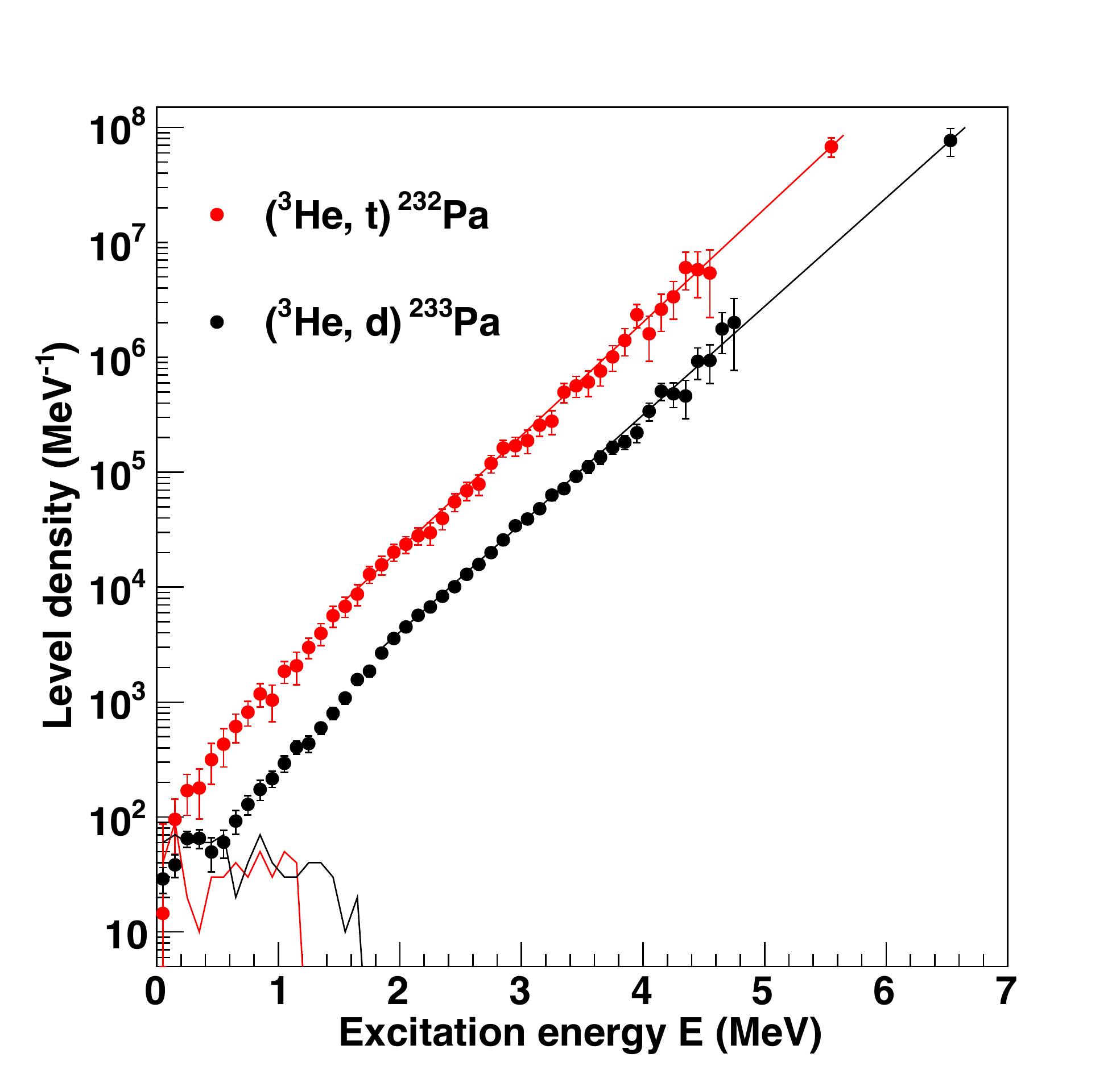}
 \caption{(Color online) Level densities for $^{232,233}$Pa. The experimental data are normalized
 to the level density of known discrete levels at low excitation energy $E$ (solid lines) and
 the level density extracted from known neutron resonance spacings $D_0$ at the neutron separation energy $S_n$.
 The connection between $\rho(S_n)$ (the upper right data points) and our experimental data 
 are made with a constant-temperature formula with $T_{\rm CT}= 0.44$ and 0.46 MeV for $^{232,233}$Pa, respectively.
 Note the extreme
 high level density for the odd-odd $^{232}$Pa, which reads $\approx$ 68 million levels per MeV at $S_n=5.55$ MeV.}
 \label{fig:232_233_pa}
 \end{center}
 \end{figure}

Fermi's golden rule predicts that the decay probability  
may be factorized into a transition matrix element between the initial and final states, 
and the density of final states~\cite{dirac,fermi}. Furthermore, according to the 
Brink hypothesis~\cite{brink}, the $\gamma$-ray transmission coefficient ${\cal {T}}$ is 
approximately independent of excitation energy. The first-generation 
matrix $P(E,E_{\gamma})$, which expresses the probability
to emit a $\gamma$-ray with energy $E_{\gamma}$ from  excitation energy $E$, 
may therefore be factorized as follows:
\begin{equation}
P(E, E_{\gamma}) \propto   {\cal{T}}(E_{\gamma}) \rho (E -E_{\gamma}),\
\label{eqn:3}
\end{equation}
where $\rho (E -E_{\gamma})$ is the level density at the 
excitation energy after the first $\gamma$-ray has been emitted in the cascades.
This factorization allows a simultaneous extraction of level density and $\gamma$-ray transmission coefficient
since the number of data points of the $P(E, E_{\gamma})$ matrix exceeds by far the number of unknown data points of the
vectors ${\cal{T}}(E_{\gamma})$ and  $\rho(E -E_{\gamma})$.
The least-square fit of ${\cal{T}}\cdot\rho$ to $P$ (see Eq.~(\ref{eqn:3}))
determines only the functional form of ${\cal{T}}$ and $\rho$.
If one solution of the functions ${\cal{T}}$ and $\rho$ is known,
one may construct identical fits to the
$P(E,E_{\gamma})$ matrix by
\begin{eqnarray}
\tilde{\rho}(E-E_\gamma)&=&A\exp[\alpha(E-E_\gamma)]\,\rho(E-E_\gamma),
\label{eq:array1}\\
\tilde{{\mathcal{T}}}(E_\gamma)&=&B\exp(\alpha E_\gamma){\mathcal{T}} (E_\gamma).
\label{eq:array2}
\end{eqnarray}
The transformation parameters $A$, $\alpha$ and $B$ can then be estimated.

    \begin{table*}[]
    \caption{Parameters used to extract level densities and $\gamma$SFs (see text).} 
    \begin{tabular}{c|c|cc|cc|cc}
    \hline
    \hline
    Reaction and                        &$S_n$   &  $\sigma(S_n)$&  $D_0$   &   $\rho(S_n)$     &  $\rho(S_n)_{\rm adopted}$    &$\langle \Gamma_{\gamma}(S_n)\rangle$  &$\langle \Gamma_{\gamma}(S_n)\rangle_{\rm adopted} $        \\
    final nucleus                       & (MeV) &               &  (eV)    &(10$^6$MeV$^{-1}$)&(10$^6$MeV$^{-1}$)&           (meV)     & (meV)              \\
    \hline
    ($^3$He,$\alpha$)\ $^{231}_{\ 90}$Th&5.118&        7.78      & 9.6(15)  & 12.7(33)          &       12.7       & 26(2)                &26\\
    ($d$, $d$')\ $^{232}_{\ 90}$Th           &6.438&        8.05      & 0.78(20)$^a$ & 30(8)$^a$     &      20          & 30(10)$^a$               &40\\
    ($^3$He,$^3$He')\ $^{232}_{\ 90}$Th &6.438&        8.05      & 0.78(20)$^a$ & 30(8)$^a$     &      30          & 30(10)$^a$               &40 \\
    ($d$, $p$)\ $^{233}_{\ 90}$Th            &4.786&        7.81      &16.5(4)  &  7.4(15)          &      4.0         & 24(2)                &20\\ \hline
    ($^3$He, $t$)\ $^{232}_{\ 91}$Pa       &5.549&        8.19      &0.51(3)   &  68(13)           &      68          & 40(1)                &35\\
    ($^3$He, $d$)\ $^{233}_{\ 91}$Pa       &6.529&        8.82      &0.42(8)   &  77(21)           &        77        & 30(10)$^a$               &45 \\     \hline
    ($d$, $t$)\ $^{237}_{\ 92}$U             &5.126&        8.02      &14.0(10)  &  9.3(19)          &       7.4        & 23(2)                &26\\
    ($d$, $d$')\ $^{238}_{\ 92}$U            &6.154&        8.26      & 3.5(8)   & 20(6)             &       20         & 30(10)$^a$                &55\\
    ($d$, $p$)\ $^{239}_{\ 92}$U             &4.806&        7.84      &20.3(6)   &  6.1(12)          &        2.45      & 23.6(8)              &33\\
    \hline
    \hline
    \end{tabular}
    \\$^a$) Estimated from systematics~\cite{RIPL3}.
    \label{tab:parameters}
    \end{table*}

\begin{table*}[]
\caption{Resonance parameters used for $\gamma$SF extrapolation.} 
\begin{tabular}{c|ccc|ccc|c|ccc|ccc}
\hline
\hline
Isotopes      &$\omega_{E1,1}$&$\sigma_{E1,1}$&$\Gamma_{E1,1}$&$\omega_{E1,2}$&$\sigma_{E1,2}$&$\Gamma_{E1,2}$&$T_f$&$\omega_{\rm pyg}$&$\sigma_{\rm pyg}$&$\Gamma_{\rm pyg}$&$\omega_{M1}$&$\sigma_{M1}$&$\Gamma_{M1}$ \\
              &   (MeV)  &      (mb)     &      (MeV)    &   (MeV)  &     (mb)      &    (MeV)      &(MeV)&    (MeV)    &      (mb)        &          (MeV)    &(MeV)  &     (mb)    &  (MeV)    \\ \hline
$^{231-233}$Th&  11.5    &      374      &       4.2     &     14.4 &      840      &     4.2       & 0.2 &      7.2    &      10          &       2.0         &  6.67 &     4.36    &    4.0   \\
$^{232,233}$Pa&  11.5    &      473      &       4.2     &     14.4 &      900      &     4.2       & 0.2 &      7.3    &      13          &       2.0         &  6.61 &     5.46    &    4.0   \\
$^{237-239}$U &  11.4    &      572      &       4.2     &     14.4 &     1040      &     4.2       & 0.2 &      7.3    &      15          &       2.0         &  6.61 &     7.00    &    4.0   \\ \hline
\hline
\end{tabular}
\\
\label{tab:GDR_param}
\end{table*}

The level density
function needs two normalization points to deduce $A$ and $\alpha$ of Eqs.~(\ref{eq:array1})
and (\ref{eq:array2}).
These points are determined at low excitation energy from the known level scheme~\cite{ENSDF},
and at high energy from the density of neutron resonances following ($n$, $\gamma$) capture
at the neutron separation energy $S_n$.
Here, the data point $\rho(S_n)$
is calculated from $\ell = 0$ neutron resonance spacings $D_0$ taken from RIPL-3~\cite{RIPL3} 
assuming the following spin distribution~\cite{GC}
\begin{equation}
g(E=S_n,I) \simeq \frac{2I+1}{2\sigma^2}\exp\left[-(I+1/2)^2/2\sigma^2\right].
\label{eq:spindist}
\end{equation}
The spin-cutoff parameter $\sigma$ at the neutron separation energy $S_n$
was estimated by use of the systematics of Ref.~\cite{egidy2}.
The values of $S_n$, $D_0$, $\sigma$ and $\rho$ are listed in Table I.
Further details on the normalization procedure
are described in Refs.~\cite{Schiller00,voin1}.

Recently~\cite{nld2013}, the level densities of $^{231-233}$Th and $^{237-239}$U were reported.
For the sake of completeness and to demonstrate the normalization procedure, we show in Fig.~\ref{fig:232_233_pa} the 
level densities for $^{232,233}$Pa. The figure demonstrates
how the level density is normalized to the anchor points at low and high excitation energies. 
It is interesting to see that only a small fraction of the levels have been 
observed in these isotopes, e.g.~at $E \approx 1 $ MeV only 10\% of all levels are known.
Above $E\approx 2$~MeV the level density follows the constant-temperature level density formula~\cite{GC},
in accordance with the findings for the other actinides. Since details on
the level densities and thermodynamics have been
presented recently~\cite{nld2013}, we will only focus on the $\gamma$SF and the appearance of the SR in the following.

\section{Normalization of the $\gamma$-ray strength function}

The actinides have a rapidly increasing level density with excitation energy due to a high density
of single-particle orbitals. Furthermore, the presence of a low pairing gap and high-$j$ orbitals surrounding
the Fermi level produce a broad spin distribution at high excitation energy. The
light-ion reactions used in this work may not populate the highest spins present in the nucleus, which in turn
could influence the shape of the observed primary $\gamma$ spectra $P$. Since the transmission coefficient
${\cal{T}}$ is assumed to be independent of spin, the observed $P$ matrix should be 
fitted with the product ${\cal{T}}\cdot\rho_{\rm red}$, where the reduced level density
is extracted by a lower value of $\rho$ at $S_n$. Since there are uncertainties
in the total $\rho(S_n)$ through the estimate of $\sigma$ and also the actual spin distribution 
brought into the nuclear system by the specific reaction, the extracted slope of ${\cal{T}}$ (the $\alpha$ parameter)
becomes rather uncertain.

The parameter $B$ controls the scaling of the transmission coefficient ${\cal {T}}(E_{\gamma})$. 
Here we use the average, total radiative width $\langle \Gamma_{\gamma} \rangle$ at $S_n$ assuming 
 that the $\gamma$-decay is dominated by dipole transitions. For initial spin $I$ and parity $\pi$, 
 the width is given by~\cite{voin1}
\begin{eqnarray}
\langle\Gamma_\gamma\rangle=\frac{1}{2\pi\rho(S_n, I, \pi)} \sum_{I_f}&&\int_0^{S_n}{\mathrm{d}}E_{\gamma} B{\mathcal{T}}(E_{\gamma})
\nonumber\\
&&\times \rho(S_n-E_{\gamma}, I_f),
\label{eq:norm}
\end{eqnarray}
where the summation and integration run over all final levels with spin $I_f$ that are accessible by $E1$ or $M1$
transitions with energy $E_{\gamma}$. 
 However,
the determination of $B$ becomes also rather uncertain because the integral of Eq.~(\ref{eq:norm})
depends on the functions of level density $\rho(E)$ and spin-cutoff parameter $\sigma(E)$.

The above complications encountered for the actinides make the standard normalization procedure of the Oslo method rather difficult to perform.
The $\alpha$ and $B$ parameters have large uncertainties, and only the $A$ parameter can be determined
with a reasonable precision. Therefore, another procedure is adopted in this work where we compare
the $\gamma$SF with the extrapolation of known data from photo-nuclear reactions.
 \begin{figure}[t]
 \begin{center}
 \includegraphics[clip,width=\columnwidth]{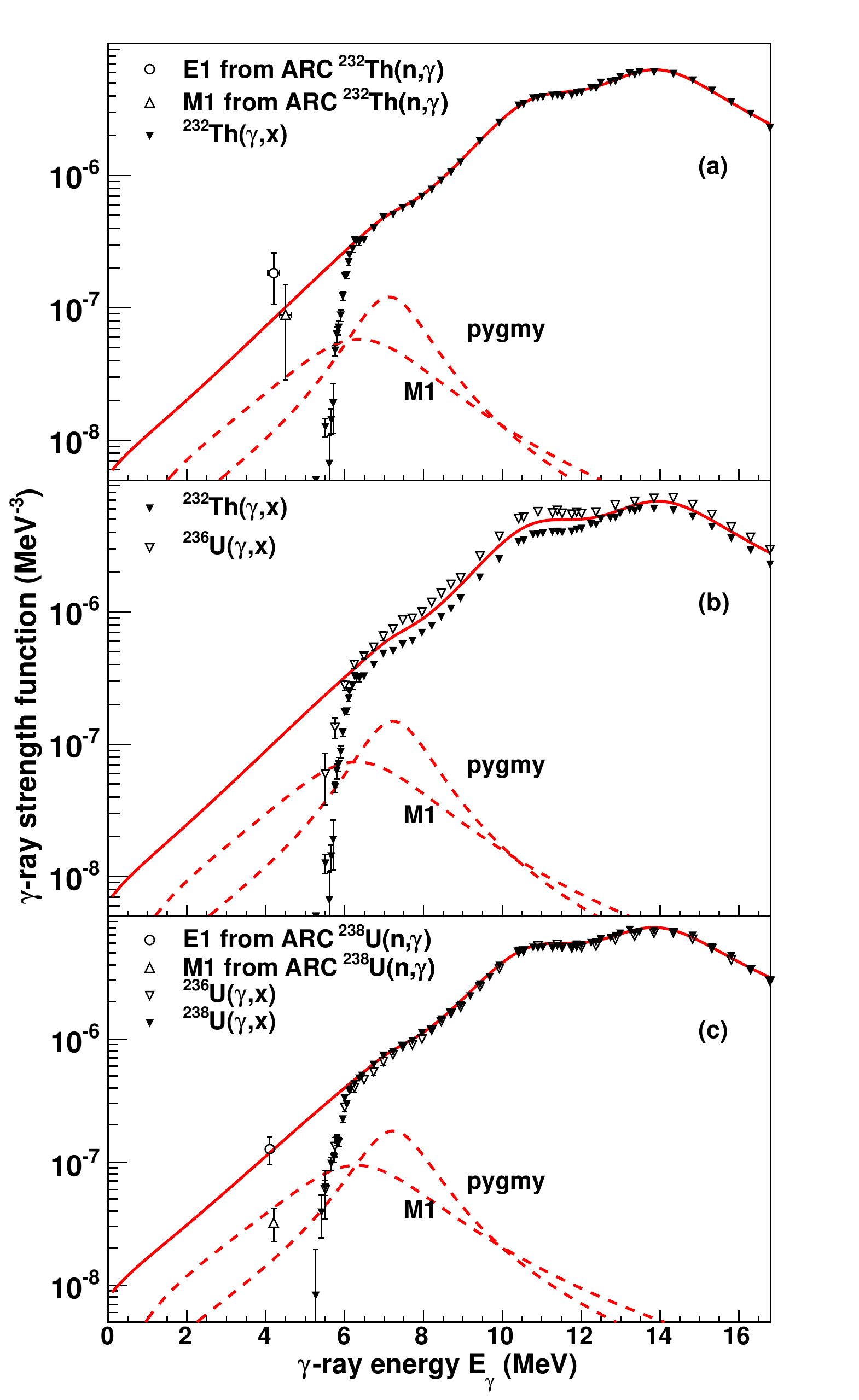}
 \caption{(Color online) Estimation of the underlying $\gamma$SF in Th (a), Pa (b) and U (c) isotopes.
 The red solid curve represents the strength expected without the scissors strength. The 
 ($\gamma$, x) data are taken from Caldwell {\em et al.}~\cite{caldwell1980} and the ARC data from Refs.~\cite{ko90,RIPL3}.
 The dashed curves are the M1 spin-flip resonance recommended by RIPL and an unknown pygmy resonance,
 which is introduced  to take into
 account the increased strength at $E_{\gamma}\approx 7.3$ MeV.}
 \label{fig:GDRs}
 \end{center}
 \end{figure}

 \begin{figure}[t]
 \begin{center}
 \includegraphics[clip,width=\columnwidth]{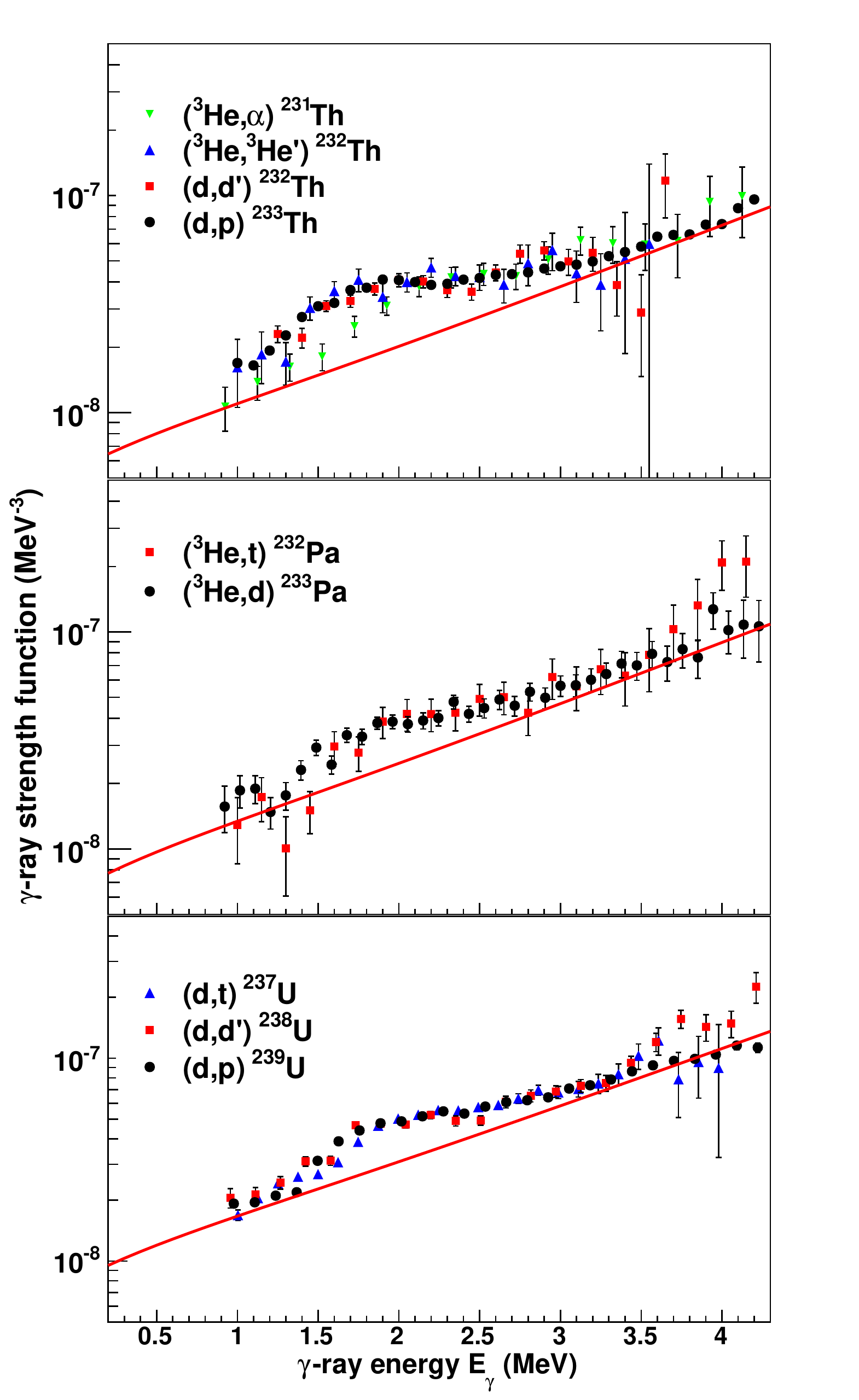}
 \caption{(Color online) Normalization of the $\gamma$-ray strength functions 
 with respect to the red solid curves of Fig.~\ref{fig:GDRs}, which are assigned to the Th, Pa and U isotopes.}
 \label{fig:norm}
 \end{center}
 \end{figure}

%
The  $\gamma$SF for dipole radiation can be calculated from the transmission coefficient ${\cal {T}}(E_{\gamma})$ by~\cite{RIPL3}
\begin{equation}
f (E_{\gamma}) =\frac{1}{2\pi} \frac{ {\mathcal{T}}(E_{\gamma})}{ E_{\gamma}^3}.
\label{eq:fT}
\end{equation}
These data are compared with the strength function derived from the cross section $\sigma$
of  photo-nuclear reactions by~\cite{RIPL3}
\begin{equation}
f (E_{\gamma}) =\frac{1}{3\pi^2 \hbar^2c^2} \frac{\sigma(E_{\gamma})}{ E_{\gamma}}.
\label{eq:fT2}
\end{equation}
In Fig.~\ref{fig:GDRs} the $\gamma$SF derived from ($\gamma$, x) cross sections on 
$^{232}$Th and $^{236,238}$U by Caldwell {\em et al.}~\cite{caldwell1980} are shown.
 Naturally, the data are seen to
drop off when $E_{\gamma}< S_n$.
Furthermore, we observe that the $\gamma$SF does not vary much with neutron number, as
seen for $^{236,238}$U in panel (c). However, a comparison between $^{232}$Th and $^{236}$U in panel (b) reveals
that the $\gamma$SF increases when the proton number goes from $Z=90$ to 92. Thus, we
assume that the $\gamma$SFs from $^{232}$Th and $^{238}$U can be applied for $^{231-233}$Th
and $^{237-239}$U, respectively. For $^{232,233}$Pa with $Z=91$, we use the average values of
$^{232}$Th ($Z=90$) and $^{236}$U ($Z=92$).

Since our data cover $\gamma$ energies below $S_n$, we have to extrapolate the ($\gamma$, x) data to lower energies
using reasonable functions.
For the double-humped giant electric dipole resonance (GEDR) we fit the data with two enhanced generalized Lorentzians
(EGLO) as defined in RIPL~\cite{RIPL3}, but with a constant temperature of the final states $T_f$. 
The ($\gamma$, x) data~\cite{caldwell1980} also reveal a resonance-like bump at
around 7.3 MeV (labeled pygmy in Fig.~\ref{fig:GDRs}). This unknown resonance\footnote{We will
not speculate here about the origin of this resonance.} together with the
$M1$ spin-flip resonance (labelled M1 in Fig.~\ref{fig:GDRs}) recommended by RIPL, are included
in the strength as standard Lorentzian shapes.
The various resonance parameters which define the solid red line shown in Fig.~\ref{fig:GDRs},
are included in Table~\ref{tab:GDR_param}. For comparison
we also include in the figure the $E1$ and $M1$ strengths derived from 
($n,\gamma$) average resonance capture data (ARC) from Ref.~\cite{ko90}.

Provided that the extrapolations in Fig.~\ref{fig:GDRs} (red solid lines) are reliable,
we may assume that this $\gamma$SF represents the "base line" with no additional strength from other resonances.
Thus, we normalize the measured $\gamma$SF to this underlying background as demonstrated in Fig.~\ref{fig:norm}. 
Here, the $\alpha$ parameter is adjusted to obtain the right slope of the observed $\gamma$SF, and $B$
is tuned to scale the data to the underlying background. 
To see the deviations from a standard normalization procedure, we also calculate
the parameter values necessary to obtain the given fit to the $\gamma$SF background. 

The adopted values for the level density and $\gamma$ width $\langle \Gamma_{\gamma}(S_n)\rangle$
 are shown in Table~\ref{tab:parameters}. In
the case of the $(d$, $p)$ reaction it seems that about half of the spin distribution at high excitation energy is
covered by the reaction. We also observe that the adopted $\langle \Gamma_{\gamma}(S_n)\rangle$ values 
deviate from the measured values. The exact reason is difficult to pin down since the
normalization integral of Eq.~(\ref{eq:norm}) depends on how the spin-cutoff parameter, level density and
transmission coefficient vary in the whole energy region up to $S_n$. The observed deviations 
may also be due to the fact that the high excitation-energy part was in some cases poorly populated 
as e.g.~for $^{232}$Th. Then the evaluation of Eq.~(\ref{eq:norm}) depends strongly on proper
extrapolations of $\rho$ and ${\mathcal{T}}$ in the unknown energy regions.

\section{The scissors resonance}
\begin{figure}[t]
 \begin{center}
 \includegraphics[clip,width=\columnwidth]{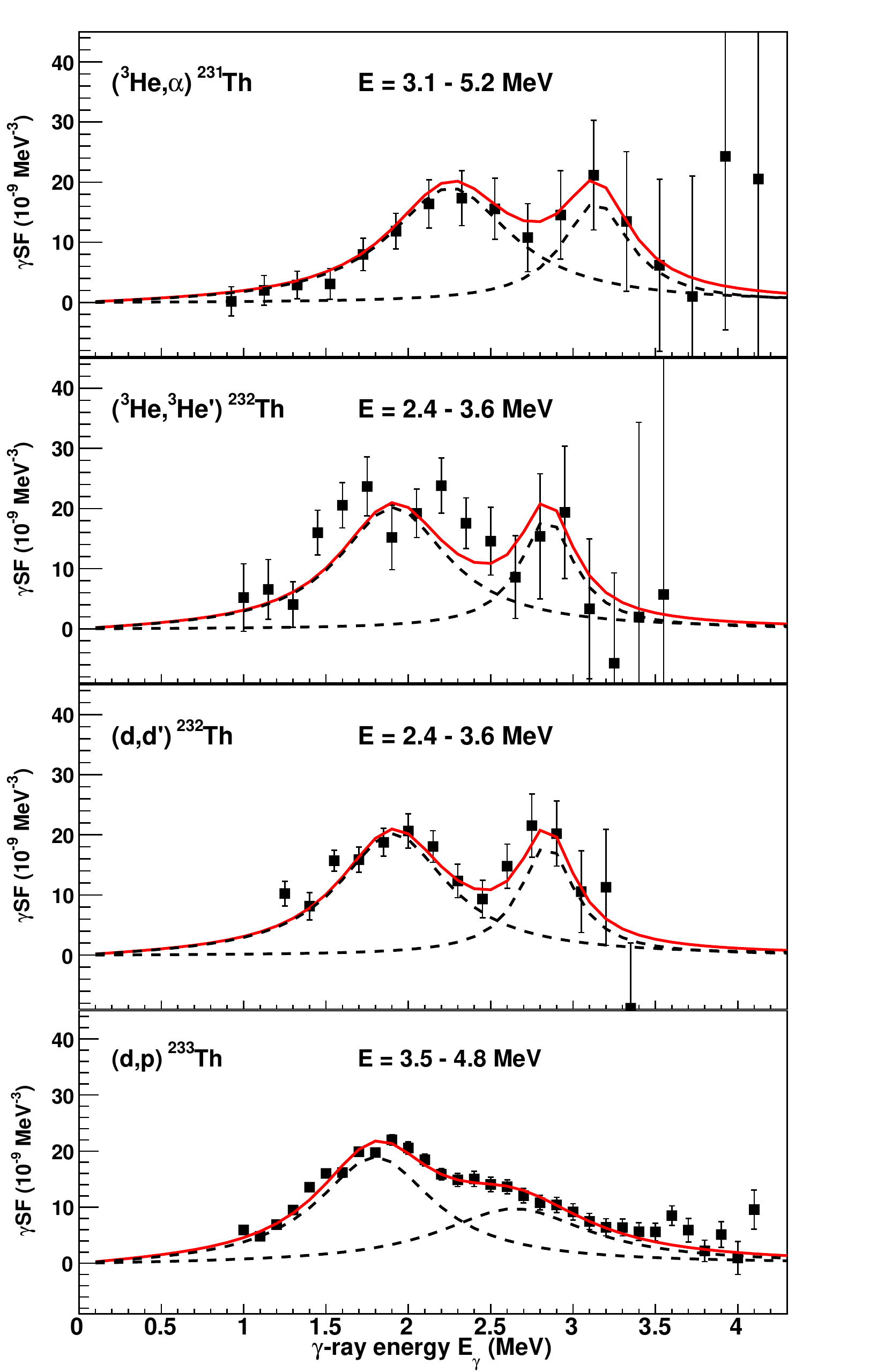}
 \caption{(Color online) The extracted scissors resonance for $^{231-233}$Th.
 The various nuclei are produced with different reactions and different
 excitation energy regions of the primary $\gamma$ matrix are utilized.}
 \label{fig:pygmy_th}
 \end{center}
 \end{figure}

\begin{figure}[t]
 \begin{center}
 \includegraphics[clip,width=\columnwidth]{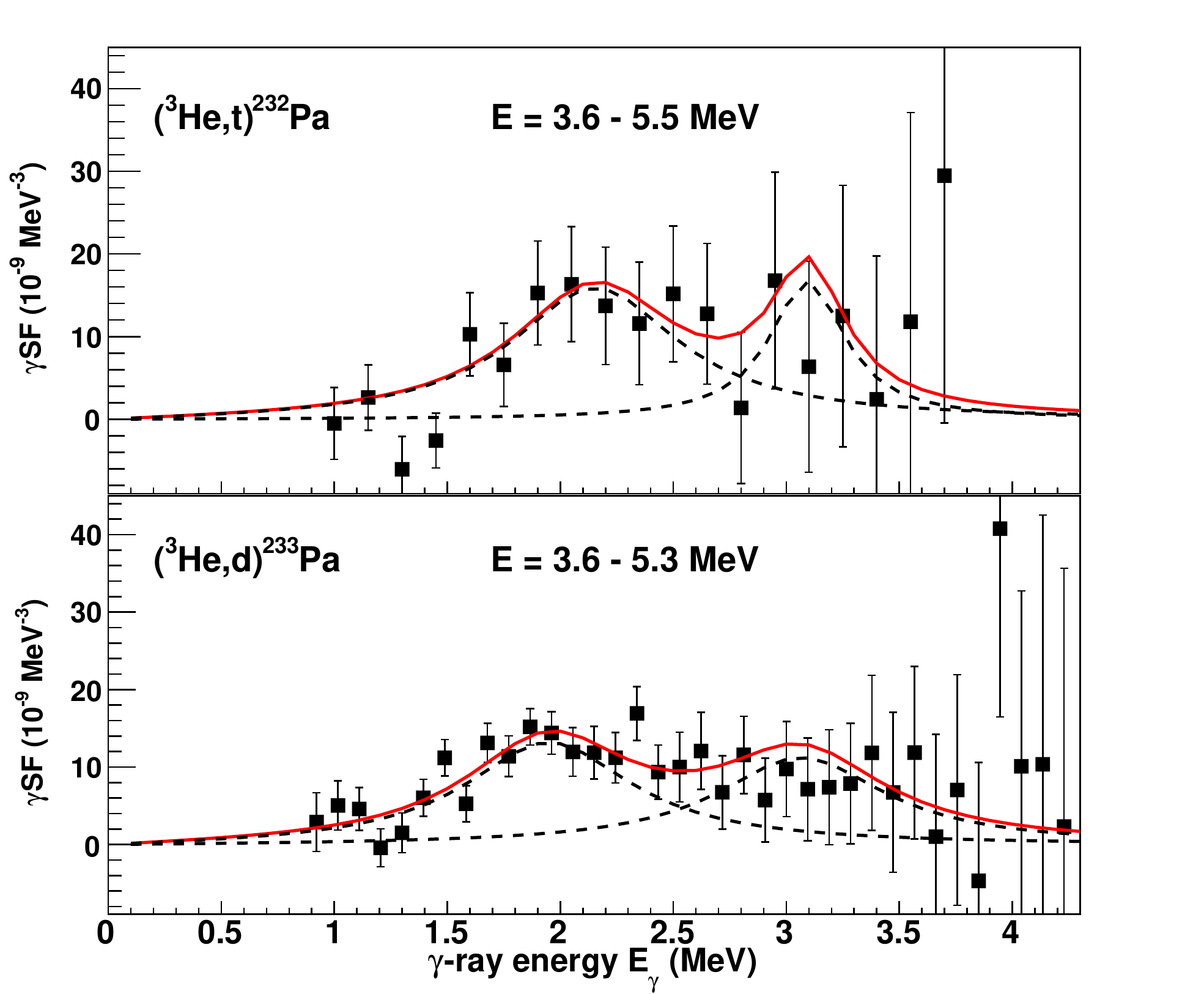}
 \caption{(Color online) Same as Fig.~\ref{fig:pygmy_th} for $^{232-233}$Pa.}

 \label{fig:pygmy_pa}
 \end{center}
 \end{figure}

\begin{figure}[h]
 \begin{center}
 \includegraphics[clip,width=\columnwidth]{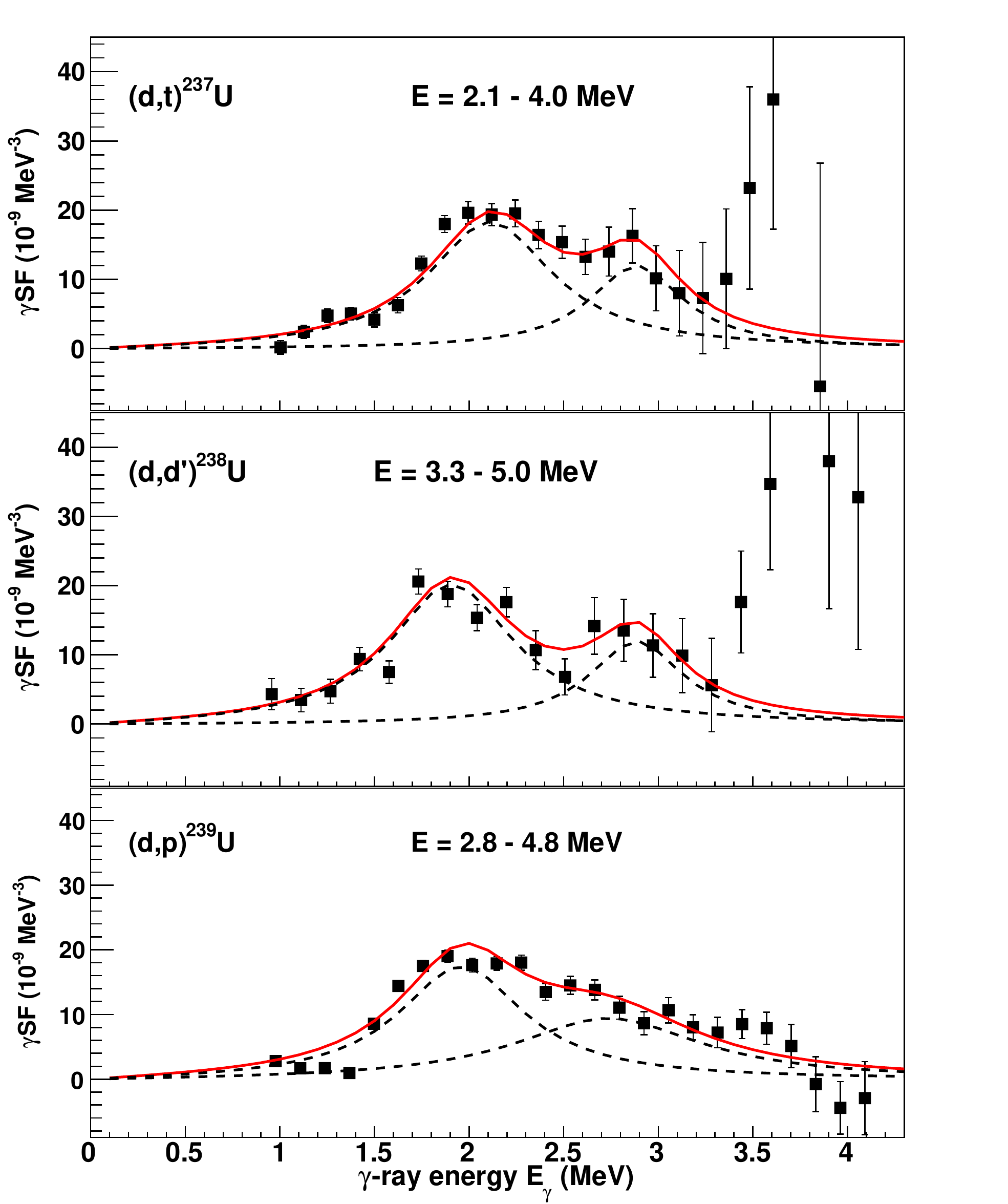}
  \caption{(Color online) Same as Fig.~\ref{fig:pygmy_th} for $^{237-239}$U.}

 \label{fig:pygmy_u}
 \end{center}
 \end{figure}

In Figs.~\ref{fig:pygmy_th}-\ref{fig:pygmy_u} we have subtracted the assumed background line of Fig.~\ref{fig:norm} 
for the thorium, protactinium and uranium isotopes. We observe a clear overshoot of strength in the
$E_\gamma=1-4$ MeV region, which is analyzed in the following. The present SR distributions differ from the ones 
previously measured~\cite{guttormsen2012}. The main reason is that 
the ($\gamma$, x) data of Gurevich {\em et al.}~\cite{gurevich} have been 
replaced by the newer and more precise data of Caldwell {\em et al.}~\cite{caldwell1980}, which gives
more reliable extrapolations, as shown in Fig.~\ref{fig:GDRs}.
Furthermore, the new NaI-response functions have slightly changed the SR $\gamma$-energy distributions.

Some of the experimental $\gamma$SFs are hampered by poor statistics, in particular for the $^{232,233}$Pa isotopes.
However, it appears that the additional $\gamma$ strength of the investigated isotopes can be decomposed into
two Lorentzians. The resonance centroid ($\omega_{{\rm SR},i}$), cross section ($\sigma_{{\rm SR},i}$)
and width ($\Gamma_{{\rm SR},i}$) are listed in Table~\ref{tab:strengths} for the lower ($i=1$) and upper ($i=2$) resonances.
From the resonance parameters, the integrated
strengths of the two components can be calculated by
\begin{equation}
    B_{{\rm SR},i}=\frac{9\hbar c}{32 \pi ^2}\left( \frac{\sigma_{{\rm SR},i} \Gamma_{{\rm SR},i}}{\omega_{{\rm SR},i}}\right).
\end{equation}
Furthermore,  the total strength and the average centroid are expressed by:
\begin{eqnarray}
B_{\rm SR}&=&\sum_{i=1,2}B_{{\rm SR},i},\\
\omega_{\rm SR} &=& \frac{\sum_{i=1,2}{ \omega_{{\rm SR},i}B_{{\rm SR},i}}}{\sum_{i=1,2}{B_{{\rm SR},i}}}.
\end{eqnarray}
In Table~\ref{tab:strengths} the upper and lower scissors strength ($B_{{\rm SR},i}$), together with
the average centroid ($\omega_{\rm SR}$) and total strength ($B_{\rm SR}$) are also listed.

Previous measurements for the SR built on the ground state~\cite{heil1988,margraf1990,yevetska2010} 
reveal centroids  around $2.2$~MeV of excitation energy, which corresponds to the first resonance in our
$\gamma$SF. Table~\ref{tab:strengths} shows that, on average, the first resonance
is centered around  $\omega_{\rm SR} = 2.05(15)$~MeV with a strength of 
$B_{\rm SR} =5.9(18)\mu_N^2 $.
Several of the mentioned studies show that levels in the $E \approx 2.2$ MeV region
have spin/parity $I^{\pi} = 1^+$, which strongly support the interpretation as the scissors resonance.
To our knowledge, the SR is the only known candidate for a soft resonance mode at these low energies.

Our data show a second component located on average $\Delta\omega_{\rm SR} = 0.89(15)$~MeV higher than the lower resonance
and with a strength of $3.5(16)\mu_N^2$.
This component was not reported in the earlier experiments~\cite{heil1988,margraf1990,yevetska2010}.
However, in a recent work~\cite{adekola2011} from the High-Intensity $\gamma$-ray Source (HI$\gamma$S) facility
at the Triangle Universities Nuclear Laboratory (TUNL)
a concentration of $1^+$ states was found at $E \approx 3$ MeV in $^{232}$Th. These data will
be compared with the present results in the next section.

\begin{table*}[]
\caption{Scissors resonance parameters and the sum-rule estimates of Eqs.~(\ref{eq:omega}) and (\ref{eq:b}) (see text).}
\begin{tabular}{c|c|cccc|cccc|cc|cc}
\hline
\hline
 Nuclide&Deformation&\multicolumn{4}{c}{Lower resonance}&\multicolumn{4}{|c}{Upper resonance}&\multicolumn{2}{|c}{Total}&\multicolumn{2}{|c}{Sum rule}\\
\hline
$^{\rm A}$X&$\delta$&$\omega_{\rm SR,1}$&$\sigma_{\rm SR,1}$&$\Gamma_{\rm SR,1}$& $B_{\rm SR,1}$ &$\omega_{\rm SR,2}$&$\sigma_{\rm SR,2}$&$\Gamma_{\rm SR,2}$& $B_{\rm SR,2}$ &$\omega_{\rm SR}$& $B_{\rm SR}$ &      $\omega_{\rm SR}$& $B_{\rm SR}$ \\
       &        &    (MeV) &(mb)    & (MeV)   &($\mu_N^2$)  &      (MeV)&(mb)     & (MeV)   &($\mu_N^2$)  &      (MeV)   &($\mu_N^2$)  &   (MeV)& ($\mu_N^2$) \\
\hline
$^{231}$Th&0.24&  2.30(15) &~0.50(5)& 0.90(10)&   6.9(11)   &  3.15(15) &0.60(20) &0.50(10) & 3.4(13)     &  2.58(15)    &  10.3(17)   &   2.0&8.6      \\
$^{232}$Th&0.24&  1.95(15) &0.45(10)& 0.80(20)&   6.5(22)   &  2.85(10) &0.60(20) &0.40(10) & 3.0(12)     &  2.23(14)    &  ~9.5(26)   &   2.0&8.6      \\
$^{233}$Th&0.24&  1.85(10) &~0.40(5)& 0.85(10)&   6.5(12)   &  2.70(20) &0.30(5)  &1.10(20) & 4.3(11)     &  2.19(15)    &  10.8(16)   &   2.0&8.5      \\ \hline
$^{232}$Pa&0.24&  2.20(20) &0.40(20)& 0.90(20)&   5.8(32)   &  3.10(30) &0.60(40) &0.40(20) & 2.7(23)     &  2.49(24)    &  ~8.5(39)   &   2.0&8.7      \\
$^{233}$Pa&0.25&  2.00(20) &0.30(20)& 0.90(30)&   4.8(36)   &  3.10(30) &0.40(30) &0.90(30) & 4.1(34)     &  2.51(25)    &  ~8.9(49)   &   2.0&9.0      \\ \hline
$^{237}$U& 0.26&  2.15(10) &~0.45(5)& 0.80(10)&   5.9(10)   &  2.90(20) &0.40(10) &0.60(15) & 2.9(11)     &  2.40(14)    &  ~8.8(15)   &   2.1&9.5      \\
$^{238}$U& 0.27&  1.95(15) &~0.45(5)& 0.80(10)&   6.5(12)   &  2.90(15) &0.40(10) &0.60(15) & 2.9(10)     &  2.24(15)    &  ~9.4(16)   &   2.2&9.8      \\
$^{239}$U& 0.25&  2.00(15) &~0.30(5)& 0.80(10)&   4.2(10)   &  2.80(15) &~0.30(5) &1.20(20) & 4.5(11)     &  2.41(15)    &  ~8.8(14)   &   2.0&9.1      \\
\hline
\hline
\end{tabular}
\\

\label{tab:strengths}
\end{table*}
\section{Comparison with other data and models}
\label{sec:comp}

\subsection{Other data}
When comparing data and model predictions for the SR built on the ground state, it is common to
quote the average excitation energy and the summed strength.
For measuring of the SR built on the ground state,
($\gamma$, ${\gamma} ^{\prime}$) and ($e, e^{\prime}$) reactions have been frequently used.
In the past, the experimental values obtained from these reactions 
were rather uncertain because many weak-intensity $\gamma$ (or $e$) lines were difficult to detect due to high backgrounds. In
addition, there were also limitations on the excitation energies covered by the experiments.
An indication of missing strength comes from the odd-mass deformed rare-earth
nuclei, which display only half of the summed strength ($\approx 1.5 \mu_N^2$)
compared to their even-even neighbors ($\approx 3 \mu_N^2$), which is rather surprising from a theoretical point of view.
The strength is fragmented into several weak and unresolved lines in the spectra 
due to 5-10 times higher level density in odd-mass nuclei.
An example is $^{163}$Dy where new and more sensitive experiments by Nord  {\em et al.}~\cite{nord03} in 2003
revealed twice the strength originally observed in 1993 by Bauske {\em et al.}~\cite{bauske93}.

For the actinides,
the second concentration of SR states at excitation energies $E \approx 2.9$ MeV was first
observed in 2011 at the HI$\gamma$S facility~\cite{adekola2011}. Prior to this study, the
second high-energy component was observed neither in
$^{232}$Th nor in $^{235,236,238}$U~\cite{heil1988,margraf1990,yevetska2010}.
The  HI$\gamma$S experiment on $^{232}$Th not only pushed the previous~\cite{heil1988} summed strength of 2.6(3) up to 4.3(6)~$\mu_N^2$,
but also revealed a two-component structure that may bring new insight to the SR mechanism.

Even though the ($\gamma$, ${\gamma} ^{\prime}$) method is based on discrete population of states
built on the ground state, a comparison  with the present results from decay in the quasi-continuum can be made. 
However, one should keep in mind that
these experiments represent two different systems with respect to the nuclear moment of inertia, as described in the following.

 \begin{figure}[t]
 \begin{center}
 \includegraphics[clip,width=\columnwidth]{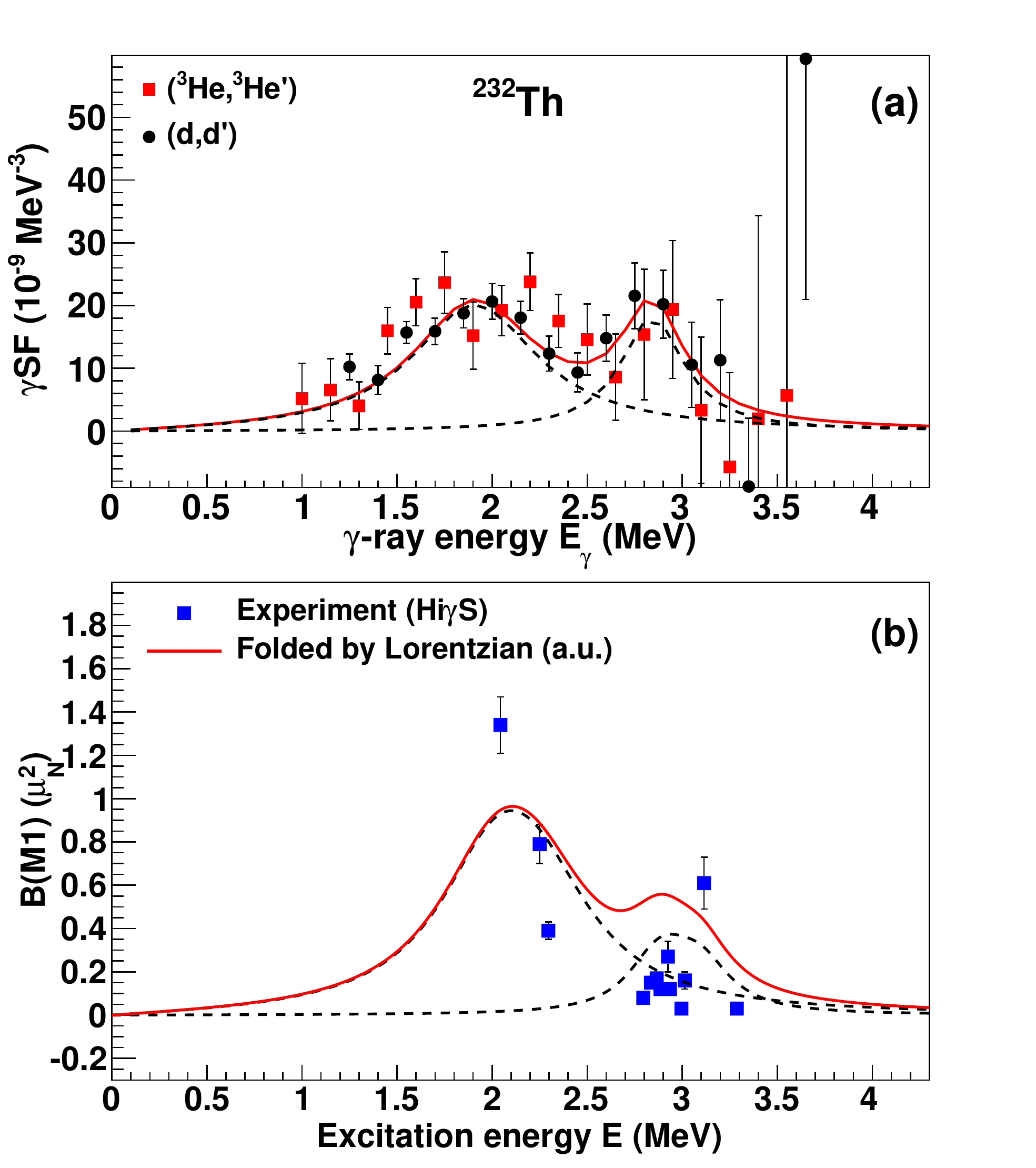}
 \caption{(Color online) Comparison between Oslo (a) and HI$\gamma$S (b) data for $^{232}$Th.
 The HI$\gamma$S data are discrete levels with measured $B(M1)$. The folded red curve of these data using Lorentzian
 shapes are shown in arbitrary units. The Oslo and  HI$\gamma$S data show some resemblance, except that the
 HI$\gamma$S data are shifted $\approx$ 300 keV up in energy and are a factor of two lower in
 summed strength than the Oslo data (see text).}
 \label{fig:nfr_oslo}
 \end{center}
 \end{figure}

Figure \ref{fig:nfr_oslo} shows the $\gamma$SF from the Oslo method (panel (a)) in $^{232}$Th
compared with the states measured at the HI$\gamma$S facility (panel (b))~\cite{adekola2011}.
One should note that the abscissa and ordinate of these plots are different. The 
HI$\gamma$S data are presented as discrete $B_{\rm SR}$ values for each state observed with
$\gamma$s of $M1$ multipolarity. According to the Brink hypothesis~\cite{brink}, these excitations should
also be built on excited states in the continuum. In order to compare with the Oslo data,
we have folded the HI$\gamma$S data using two Lorentzians with widths of $\Gamma=0.75$ and
0.35 MeV for states below and above excitation energy $E=2.5$ MeV, respectively. These widths are
chosen somewhat smaller than the widths extracted from the Oslo data in Table~~\ref{tab:strengths} since the spread in
the energy positions of the $1^{+}$ states also
contributes to the width. We see from Fig.~\ref{fig:nfr_oslo}~(b)
that the two resonance peaks are located $\approx 300$ keV higher than for the Oslo data.
The total strength measured by the HI$\gamma$S group~\cite{adekola2011} is $4.3(6)\mu_N^2$
versus the higher value of $9.5(26)\mu_N^2$ in the present study.

These results are not necessarily representing a controversy.
Similar deviations have been found for the scissors strength in the deformed rare earth region
where ($\gamma$, ${\gamma} ^{\prime}$) experiments~\cite{heyde2011} 
typically yield strengths of $B_{\rm SR} = 3 - 4\ \mu_N^2$.
Various measurements of the $\gamma$-decay between levels 
in the quasi-continuum show significant higher SR strength.
Here, the two-step cascade method and the Oslo method 
give integrated strengths of $6-7\ \mu_N^2$~\cite{milan2004,schi2006}.
One could speculate if the lower strength in ($\gamma$, ${\gamma} ^{\prime}$)
experiments is due to missing states caused by low $\gamma$ intensities relative to the
background or because of limited excitation-energy regions.
However, the deviation may also be due to the fact that the 
scissors strength depends on the moment of inertia that takes
different values for the ground state and the levels in the quasi-continuum.

From theoretical considerations described in the next section, the strength of the SR should be proportional to the moment of inertia,
which may take a lower and upper limit. In principle, for the SR built on the ground
state, the ground-state moment of inertia should be applied.
This quantity is easily extracted from
the first rotational $2^+$ state in even-even deformed nuclei by
\begin{equation}
\Theta_{\rm gs}=3\hbar ^2/E_{2^+}.
\end{equation}
For the SR in the quasi-continuum, the rigid-body moment of inertia should be used:
\begin{equation}
    \Theta_{\rm rigid} =\frac{2}{5}m_N r_0^2 A^{5/3}(1+0.31\delta),
    \label{eq:theta}
\end{equation}
    with $r_0=1.15$ fm and $\delta$ is the nuclear quadrupole 
    deformation\footnote{In this work, we use the  quadrupole deformation parameter $\delta$, which
    relates to the deformation parameters $\epsilon_2$ and $\beta_2$; to lowest order
    $\delta \approx \epsilon_2 \approx \beta_2\sqrt{45/16\pi}$.} taken from~\cite{goriely2009}.

In the case of $^{232}$Th, we have $E_{2^+}=0.0494$ MeV giving the lower limit $\Theta_{\rm gs}/\hbar^2 = 60.7$~MeV$^{-1}$,
while the rigid value becomes $\Theta_{\rm rigid}/\hbar^2 = 120.8$~MeV$^{-1}$, which represents the upper limit.
It is interesting that the ratio $\Theta_{\rm rigid}/\Theta_{\rm gs} = 2.0$ is in agreement
with the ratio $\sum B(M1)_{\rm Oslo}/\sum B(M1)_{{\rm HI}\gamma{\rm S}}=2.2(7)$ for $^{232}$Th.
A similar scaling is valid also for the well-deformed rare-earth region.
These observations may call for a consistent model that is capable of describing the
SR states built on the ground state as well as the SR distribution in the quasi-continuum.

\subsection{Models}

Numerous SR models have been launched to explain the results
of the ($\gamma$, ${\gamma} ^{\prime}$) and ($e, e^{\prime}$) reactions~\cite{heyde2011}.
The predictions for deformed rare-earth nuclei were often guided by
the measured values found at the time when the models were published.
Quasiparticle Random Phase Approximation (QRPA) models are rather popular, although
these also have some freedom for tuning the results to experimental data.
A common definition~\cite{nojarov93} of an SR state is
when the orbit-to-spin ratio is $|M_l/M_s|^2 \gg 1$.
In the work of Kuliev {\em et al.}~\cite{kuliev10},
QRPA calculations were performed for the $E=2-4$~MeV excitation region. Their
calculations for $^{232}$Th and $^{236,238}$U give typical strengths of
$\sum B = 5-6 \mu_N^2$ at the average excitations energy of $E=2.6$ MeV.
With a moment of inertia ratio of $\Theta_{\rm rigid}/\Theta_{\rm gs} \approx 2.0$
these predictions are in agreement with the present findings.

It is interesting to investigate the most important single-particle orbitals responsible
for the SR in the QRPA calculations~\cite{kuliev10}. The various SR states are composed
of several pairs of Nilsson orbitals, having $\Delta\Omega=1$.
The most pronounced pairs of the strongest SR states at low excitation 
in $^{232}$Th and $^{238}$U are $\frac{1}{2}^-[530]_{p}\otimes\frac{3}{2}^-[521]_{p}$
and $\frac{5}{2}^+[642]_{p}\otimes\frac{7}{2}^+[633]_{p}$, respectively. 
The strongest and higher-lying SR states of $^{232}$Th, are calculated to
have excitation energies of 2.998 and 3.134 MeV. Their wave-functions are dominated by~\cite{kuliev2013} the
$\frac{3}{2}^+[402]_{p}\otimes\frac{5}{2}^+[402]_{p}$
and $\frac{1}{2}^+[541]_{p}\otimes\frac{1}{2}^+[530]_{p}$ configurations, respectively.
The mechanism behind the splitting of the strength into two energy regions is not clear, 
other than the distance of the Nilsson orbitals to the
Fermi surface has some relevance. The strong admixture of 
many two-quasiparticle orbitals in the SR states
indicates that these excitations are rooted in collective motion.

In this work we have chosen the sum-rule approach~\cite{lipparini1989}, which 
is a rather fundamental way  of  predicting both $\omega_{\rm SR}$ and $B_{\rm SR}$.
The drawback is that only these two gross properties are given. This approach
requires that the strength is located at one specific excitation energy, and is not able to explain why
the SR distribution splits into two components. 

We follow
the description of Enders {\em et al.}~\cite{enders2005}
with the exception that the ground-state moment of inertia
will be replaced by the rigid-body moment of inertia.
The inversely and linearly energy-weighted sum rules are given by 
\begin{eqnarray}
S_{+1}&=&\frac{3}{8\pi}\Theta_{\rm rigid}\delta^2\omega_D^2(g_p-g_n)^2 ~\left[ \mu^2_N {\rm MeV} \right], \\
S_{-1}&=&\frac{3}{16\pi}\Theta_{\rm IV}(g_p-g_n)^2 ~\left[ \mu^2_N {\rm MeV}^{-1} \right].
\end{eqnarray}
For the $g$ factors\footnote{Bare gyromagnetic factors are $g_p=1$ and $g_n=0$.}
we use the common expression $g_p - g_n\approx 2Z/A$,
which rests on the assumption that the neutron and rotational gyromagnetic factors are $g_n \approx 0$
and $g_R \approx (g_p + g_n)/2 \approx Z/A$, respectively~\cite{BM75}.
   Since the SR is measured in the quasi-continuum, the isovector moment 
   of inertia $\Theta_{\rm IV}$ is taken as the rigid-body moment of inertia 
    $\Theta_{\rm rigid}$ as discussed above.
    
According to Enders {\em et al}.~\cite{enders2005} the $K=1$ component of the
isovector giant quadrupole resonance (IVGQR) will dominate $S_{+1}$
and has to be removed using a reduction factor
\begin{equation}
\xi=\frac{\omega_Q^2}{\omega_Q^2 + 2\omega_D^2}
\end{equation}
that depends on the energy centroids of the isovector giant dipole (IVGDR)
and isoscalar giant quadrupole (ISGQR) resonances:
\begin{eqnarray}
\omega_D &\approx& (31.2A^{-1/3} + 20.6A^{-1/6})(1-0.61\delta) {\rm MeV}, \\
\omega_Q &\approx& 64.7A^{-1/3}(1-0.3\delta) {\rm MeV}.
\end{eqnarray}
In the mass region investigated here, $\xi$ is rather independent on $A$ (and $\delta$)
and takes the value $\xi \approx 0.27$.
The adequate expression of $S_{+1}$ for the low-lying scissors mode then reads:
\begin{equation}
S_{+1}=\frac{3}{2\pi}\Theta_{\rm rigid}\delta^2\omega_D^2g_{\rm IS}^2\xi, \\
\end{equation}
where $g_{\rm IS}=\frac{1}{2}(g_p + g_n)\approx Z/A$.

The two sum rules can now be utilized to extract the SR centroid and strength:
\begin{eqnarray}
\omega_{\rm SR}&=& \sqrt{S_{+1}/S_{-1}} \nonumber  \\
       &=&\delta \omega_D\sqrt{2\xi}, \label{eq:omega} \\
B_{\rm SR}&=& \sqrt{S_{+1}S_{-1}} \nonumber \\
          &=&\frac{3}{4\pi}\left(\frac{Z}{A}\right)^2 \Theta_{\rm rigid}\delta \omega_D\sqrt{2\xi}.
          \label{eq:b}
\end{eqnarray}
It is interesting to note that $S_{-1}$ does not depend on $\xi$. Thus, if the experimental $\omega_{\rm SR}^{\rm exp}$ is known,
a less rigorous relation for the strength is:
\begin{eqnarray}
    B_{\rm SR}&=&\omega_{\rm SR}^{\rm exp}S_{-1} \nonumber \\
              &=& \omega_{\rm SR}^{\rm exp}\frac{3}{4\pi}\left(\frac{Z}{A}\right)^2 \Theta_{\rm rigid},
\end{eqnarray}
which replaces the centroid from the sum rule with the experimental value. However, this
was not necessary in the present work since both the centroid and strength are well described by the sum rules.

The total SR strength and weighted centroid for the eight nuclei of Figs.~\ref{fig:pygmy_th}-\ref{fig:pygmy_u} 
are summarized in Table~\ref{tab:strengths}. 
The two last columns of Table~\ref{tab:strengths} show the predicted sum-rule estimates.
Both the $\omega_{\rm SR}$ and $B_{\rm SR}$ values are in good agreement with our measurements.
Although $S_{-1}$ depends very weakly on $\delta$, the $\omega_{\rm SR}$ follows a $\delta$ dependence, see Eq.~(\ref{eq:omega}).
However, since $B_{\rm SR}=\omega_{\rm SR} S_{-1}$ the strength follows $\approx \delta$,
contrary to the strong $\delta^2$ dependence for SR states built on the ground state~\cite{ziegler1990,neumann1995}.
Unfortunately, our data do not allow us to conclude any systematic behavior regarding  $A$ or $\delta$;
all eight nuclei display the same resonance parameters within the experimental errors.
 \begin{figure}[t]
 \begin{center}
 \includegraphics[clip,width=\columnwidth]{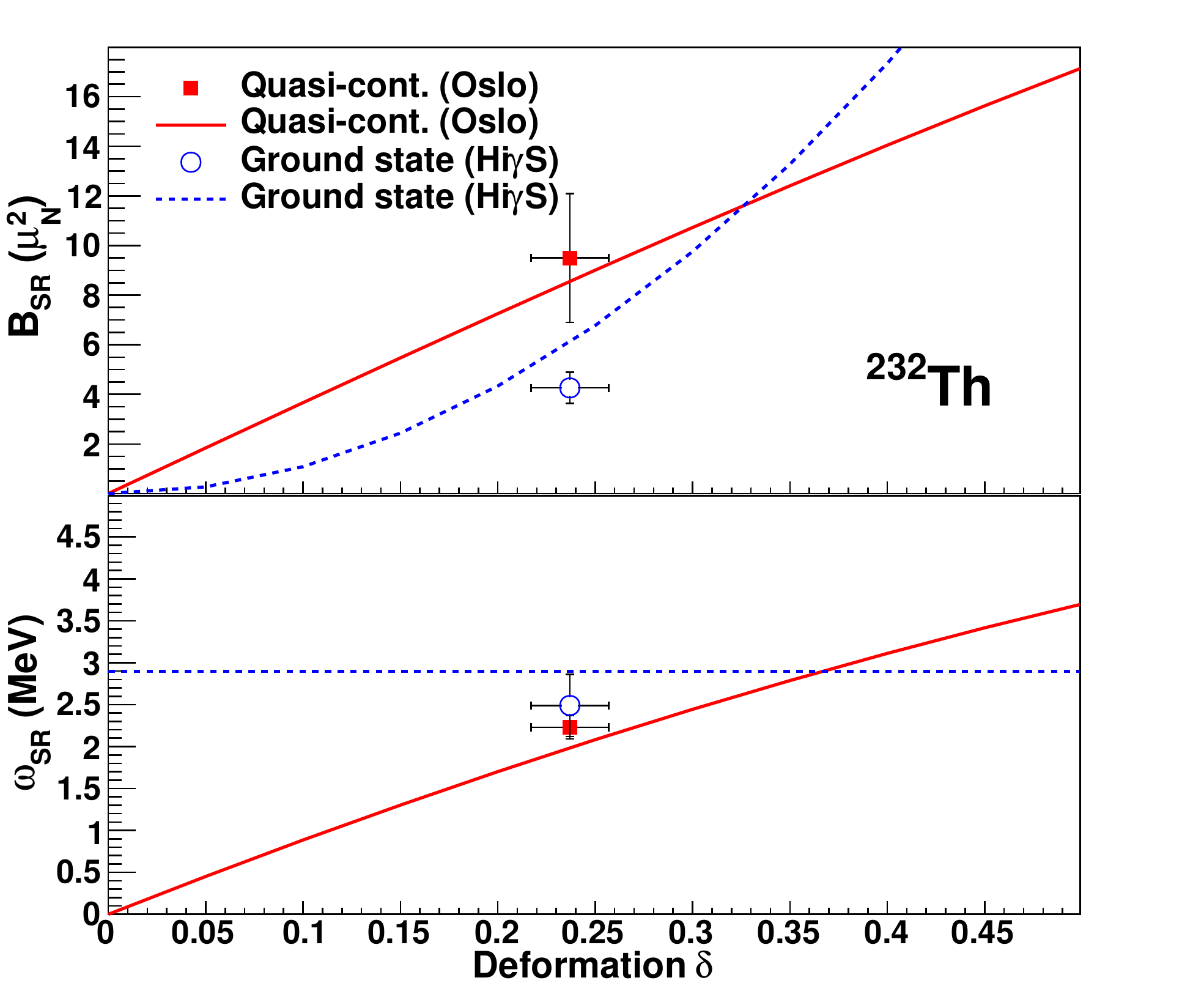}
 \caption{(Color online) Comparison between observed SRs for $^{232}$Th from Oslo and ${\rm HI}\gamma{\rm S}$.
 The data are compared with sum-rule estimates using $\Theta_{\rm rigid}$ and $\Theta_{\rm gs}$ for the
 two experiments, respectively. The deformation dependence for SR built on the ground state is assumed
 to be constant for $\omega$ and follows a $\delta^2$-rule for $\sum B(M1)$.}
 \label{fig:sumrule}
 \end{center}
 \end{figure}

In Fig.~\ref{fig:sumrule} we have plotted the sum-rule estimates for $^{232}$Th and compared to the experimental
values. With the assumption of a rigid-body moment of inertia in the quasi-continuum, the Oslo data
are very well reproduced at a deformation of $\delta=0.24$. For illustration, it is interesting to
show the sum-rule estimates for different deformations, still assuming the $^{232}$Th system. The sum rule predicts that
the centroid as well as the strength will decrease linearly with $\delta$ as one approaches more spherical nuclei.
For the HI$\gamma$S data the strength and
centroid are overestimated by the sum-rule approach using ground-state moment of inertia. In this case, one cannot
calculate the $\delta$ dependence directly from the sum rule as the $E_{2^+}$ energy is unknown for deformations
differing from the deformation of $^{232}$Th with $\delta\approx 0.24$.
However, it is well known
that the average centroid is approximately constant and the strength
follows a $\approx \delta^2$ rule~\cite{enders2005}. These dependencies are indicated as dashed blue lines
in Fig.~\ref{fig:sumrule}. It would be very interesting to follow the SR in the quasi-continuum
to lower deformations to see if the strength and centroid decrease as expected.

Over the last 30 years many theoretical works have been published for the SR built on the 
ground state~\cite{heyde2011}. However, the centroid and strength of the SR in the quasi-continuum is the
quantity that directly relates to the reaction rates in e.g.~astrophysical environments. For example for the r-process,
which involves nuclei with extreme $N/Z$ ratios, 
the decrease in neutron-separation energy with neutron number is expected to give an
increasing impact from the SR on the reaction rates.
The SR represents also an important ingredient for the simulations of fuel cycles for
fast nuclear-power reactors. Sensitivity and uncertainty studies~\cite{aliberti2006,aliberti2004} 
for reactors included in
the Generation IV (Gen IV) initiative and Accelerator Driven Systems (ADS)
show that the cross sections involved must be known with high precision.
Thus, there is a great need for new theoretical and experimental
investigations of the summed SR strength, its dependence on the deformation, and the origin of the
two-component structure seen here in the quasi-continuum of the actinides.

\section{Conclusions}
\label{sec:con}

The level densities of $^{232-233}$Pa and the $\gamma$SFs of $^{231-233}$Th, $^{232,233}$Pa 
and $^{237-239}$U have been determined using the Oslo method. The level densities show
a constant-temperature behavior as recently reported for $^{231-233}$Th and $^{237-239}$U.

All the eight actinides investigated show an excess in the $\gamma$SFs in the $E_{\gamma} = 1 - 4$~MeV region,
which is interpreted as the scissors resonance in the quasi-continuum. The underlying strength has been subtracted
by extrapolating the
assumed strength from the tails of other resonances; the double humped GEDR, the spin-flip GMDR and an unknown
pygmy resonance. 

The sum-rule applied to the quasi-continuum gives a satisfactory description of the
SR for all isotopes studied. The approach predicts that $\omega_{\rm SR}$ and $B_{\rm SR}$ are proportional to
the deformation parameter $\delta$. This is in contrast with the $\delta^2$ behavior of the SR built on the ground state.
Furthermore, the SR shows a splitting into two components, which is in accordance with data
 from the HI$\gamma$S facility. However, there are currently no firm theoretical
explanations of the two-component structure seen in the present study. 
Theoretical and experimental studies of the SR in the quasi-continuum
are called for to obtain reliable
reaction rate predictions used in nuclear-astrophysics and reactor applications.

\acknowledgements
We would like to thank J.C.~M{\"{u}}ller, E.A.~Olsen, A.~Semchenkov and J.~Wikne at the 
Oslo Cyclotron Laboratory for providing the stable and high-quality deuterium and 
$^3$He beams during the experiment, the Lawrence Livermore National Laboratory for
providing the $^{232}$Th target and the GSI Target Laboratory for the production 
of the $^{238}$U target. 
We thank A.A.~Kuliev, E.~Guliyev and F.~Ertugral for sharing details of their QRPA calculations.
This work was supported by the Research Council of Norway (NFR),
the French national research programme GEDEPEON, the US Department of Energy 
under Contract No.~DE-AC52-07NA27344, the National Research Foundation of South Africa, 
the European Commission within the 7th Framework Programme through 
Fission-2010-ERINDA (Project No.~269499) and by the 
European Atomic Energy CommunityÕs 7th Framework Programme under grant agreement no.~FP7-249671 (ANDES).

\vfill
\end{document}